\documentclass[USenglish,oneside,twocolumn]{article}

\usepackage[utf8]{inputenc}
\usepackage[big]{dgruyter_NEW}
\usepackage{color}
\usepackage{placeins}
\usepackage{afterpage}
\usepackage{float}

\newcolumntype{L}[1]{>{\raggedright\let\newline\\\arraybackslash\hspace{0pt}}m{#1}}
\newcolumntype{$}{>{\global\let\currentrowstyle\relax}}
\newcolumntype{^}{>{\currentrowstyle}}

\begin{document}

  \author*[1]{Noah Apthorpe}

  \author[2]{Dillon Reisman}

  \author[3]{Srikanth Sundaresan}

  \author[4]{Arvind Narayanan}
  
  \author[5]{Nick Feamster}

  \affil[1]{Princeton University, E-mail: apthorpe@cs.princeton.edu}

  \affil[2]{Princeton University, E-mail: dreisman@princeton.edu}

  \affil[3]{Princeton University, E-mail: srikanths@princeton.edu}

  \affil[4]{Princeton University, E-mail: arvindn@cs.princeton.edu}
  
  \affil[5]{Princeton University, E-mail: feamster@cs.princeton.edu}

  \title{\huge Spying on the Smart Home: Privacy Attacks and Defenses on Encrypted IoT Traffic}

  \runningtitle{Spying on the Smart Home: Privacy Attacks and Defenses on Encrypted IoT Traffic}

  \begin{abstract}
{The growing market for smart home IoT devices promises new conveniences for consumers while presenting new challenges for preserving privacy within the home.  Many smart home devices have always-on sensors that capture users' offline activities in their living spaces and transmit information about these activities on the Internet.  
In this paper, we demonstrate that an ISP or other network observer can infer privacy sensitive in-home activities 
by analyzing Internet traffic from
smart homes containing commercially-available IoT devices \textit{even when the devices use encryption}. We evaluate several strategies for mitigating the privacy risks associated with smart home device traffic, including blocking, tunneling, and rate-shaping. Our experiments show that traffic shaping can effectively and practically mitigate many privacy risks associated with smart home IoT devices. We find that 40KB/s extra bandwidth usage is enough to protect user activities from a passive network adversary. This bandwidth cost is well within the Internet speed limits and data caps for many smart homes.  
}
\end{abstract}
  \keywords{Privacy, Internet-of-things, smart homes, traffic analysis}
  \startpage{1}

\maketitle

\section{Introduction}
\label{sec:intro}

Internet-connected physical devices have rapidly increased in popularity and commercial availability within the past several years. 
This trend, the {\em Internet of Things (IoT)}, includes many consumer products that often replace conventional non-networked home appliances or introduce new technologies into consumer homes. 
Unfortunately, a future where ``smart'' homes are filled with numerous Internet-connected devices can also introduce significant privacy concerns.

\begin{figure}[t]
\centering
\includegraphics[width=0.49\textwidth]{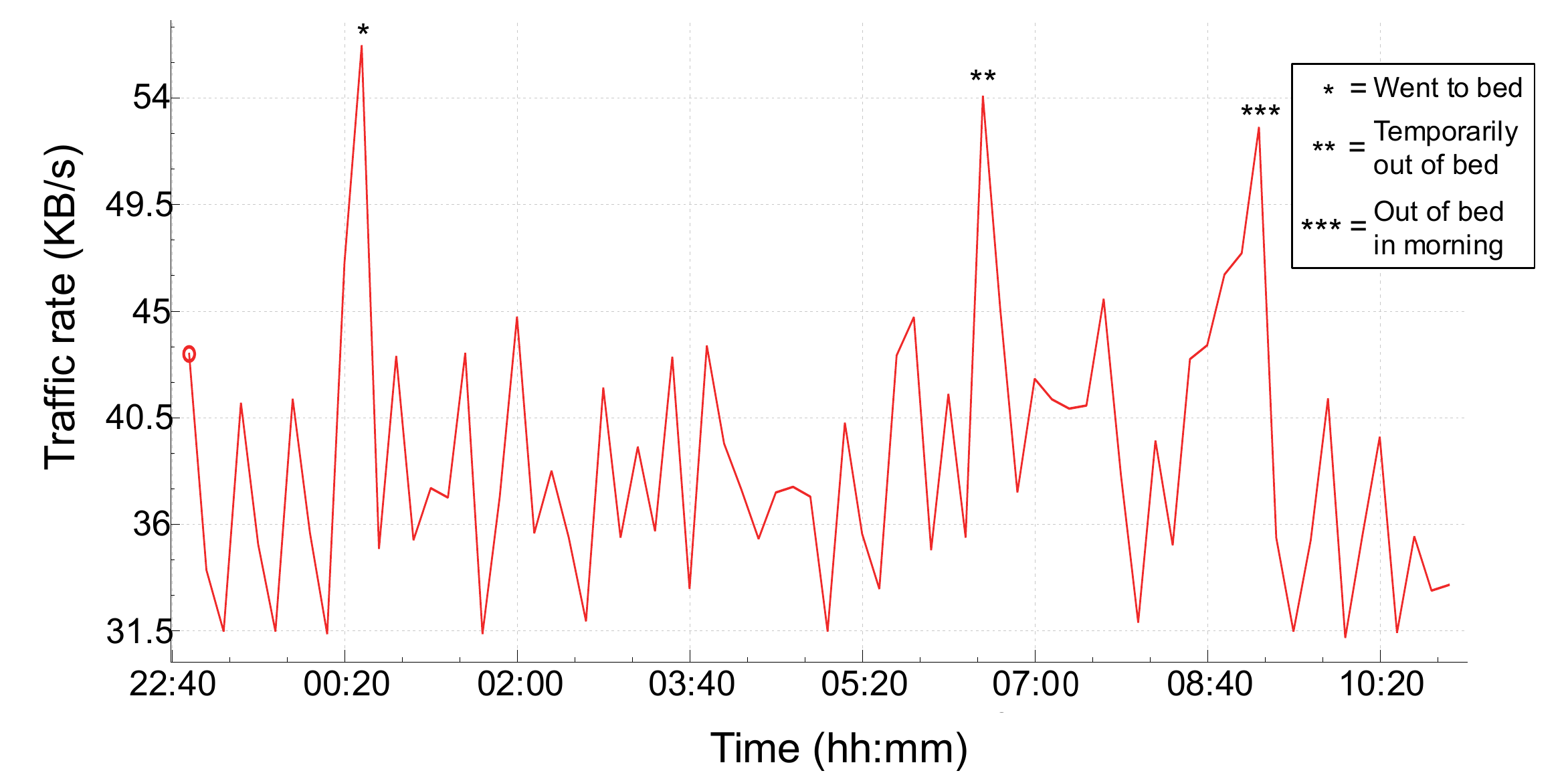}
\caption{Traffic rate to and from a Sense Sleep Monitor over a 12-hour period.  A network observer could easily determine a user's sleeping habits without needing access to encrypted traffic content. }
\label{fig:sleep-pcap}
\end{figure}

Many smart home devices have always-on sensors that capture users' offline activities in their living spaces and transmit information about these activities outside of the home, typically to cloud services run by device manufacturers. 
Examples of offline activities recorded by currently available smart home devices include sleeping patterns~\cite{sense, beddit}, exercise routines~\cite{iot-exercise}, child behaviors~\cite{iot-bears}, medical information~\cite{iot-medical}, and sexual activity~\cite{iot-sex}.
Even if a smart home device is not designed to capture privacy sensitive activities, such activities may indirectly influence information collected by device sensors, allowing them to be identified by inference techniques.  

Consumers are generally concerned about the privacy implications of smart home devices.  
Surveys have shown that many individuals do not want specific in-home behaviors to be recorded or shared by IoT devices~\cite{choe2011living}. 
While purchasers of smart home devices might expect that the device manufacturer can track how a consumer uses its device, smart home network traffic is susceptible to eavesdropping by other parties. 
Such parties include Internet service providers (ISPs), Wi-Fi eavesdroppers, or state-level surveillance entities. 

The privacy threat posed by ISPs is especially concerning given the recent reversal of the FCC's broadband consumer privacy rules~\cite{fcc-reversal}. These rules, approved in October 2016 but never enacted, placed restrictions on ISPs' ability to collect and use consumer information gleaned from traffic analysis or deep packet inspection~\cite{fcc-pr}.  In the debate surrounding these regulations, some argued that the privacy risks of smart home device traffic analysis are minimal, because the increasing pervasiveness of encryption prevents network observers from viewing sensitive data in traffic content~\cite{swire}. 
In this paper, we demonstrate that despite the broad adoption of transport layer encryption, smart home traffic \emph{metadata} is sufficient for a passive network adversary to infer sensitive in-home activities (Figure~\ref{fig:sleep-pcap}).
We present an attack on user privacy using metadata from these devices that is effective \textbf{even when devices use encryption}. The attack requires two steps:

\begin{enumerate}
\item {\em Use Domain Name System (DNS) queries or device fingerprinting to identify smart home devices from network traffic.}   This alone can be a privacy risk. For example, learning that a someone owns an IoT blood sugar monitor or pacemaker effectively reveals a diabetes or heart-disease diagnosis, respectively.

\item \textit{Infer user activities from changes in device traffic rates.} Once an adversary identifies a device and knows its purpose, device states, reflected in traffic rates, directly imply user behaviors. 
\end{enumerate}

\noindent
We tested this attack on several commercially available smart home devices and found that all revealed potentially private user behaviors through network traffic metadata. Traffic rates from a Sense sleep monitor revealed consumer sleep patterns, traffic rates from a Belkin WeMo switch revealed when a physical appliance in a smart home is turned on or off, and traffic rates from a Nest Cam Indoor security camera revealed when a user is actively monitoring the camera feed or when the camera detects motion in a user's home.  The general effectiveness of this attack across smart home device types and manufacturers motivates the need for technical privacy protection strategies.  

In testing smart home devices, we discover that many devices do not work without an active Internet connection.
This means that firewalling smart home devices from the Internet is not an effective means of mitigating the problem of traffic rate metadata.
Additionally, tunneling smart home traffic through a VPN makes the traffic metadata privacy attack considerably more challenging, but does not provide guaranteed protection.
We find that certain common device combinations and user activity patterns minimize the ability of a VPN to obfuscate smart home traffic metadata.  

We demonstrate that traffic shaping by independent link padding (ILP) provably prevents the metadata attack while preserving device functionality.  
ILP involves shaping traffic rates to match a predetermined rate or schedule, thereby exposing no information about device behavior to an adversary. 
Implementations typically involve padding or fragmenting all packets to a constant size and buffering device traffic or sending cover traffic to enforce the predetermined rate.
We empirically determined shaped traffic rates resulting in low bandwidth overheads and tolerable network latencies for a variety of smart home devices.  

For smart homes without devices that stream audio or video, we found that only a maximum of 7.5 KB/s of cover traffic is needed to completely mask user activities. This is $<0.4\%$ of the 2016 average broadband upload and download speeds in the United States \cite{speedtest}. This rate of cover traffic would result in approximately 19GB of extra data used per month, much less than the 1TB monthly broadband data caps enforced by the largest United States and Canadian ISPs \cite{comcast-data-cap, bell-internet}.

For smart homes with devices that stream audio or video, we found that no more than 40KB/s of cover traffic per household was needed for effective ILP shaping. While more costly than shaping for non-A/V devices, it is still $<2\%$ of the average upload speeds of United States broadband internet. 

Although ILP shaping is well-understood, it is typically dismissed as requiring excessive latency or bandwidth overhead to be practical for real-world use. 
Our results contradict this common belief. 
ILP traffic shaping is a reasonable privacy protection method for smart homes with rate-limited broadband access or data caps. 
\section{Related Work}
\label{sec:related}

This paper draws on a rich history of related research on traffic analysis attacks and prevention techniques.  The attack we describe is similar in spirit to the Fingerprint and Timing-based Snooping (FATS) attack presented by Srinivasan et al.~\cite{srinivasan2008protecting}. The FATS attack involves  activity detection, room classification, sensor classification, and activity recognition from Wi-Fi traffic metadata from a sensor network deployed in the home---the precursor to today's smart home IoT devices. In contrast to the attacks that we present, the FATS attack relies on wireless network traffic instead of a observations from a last-mile Internet service provider or other adversary located on the WAN. This context allows techniques such as radio fingerprinting and signal attenuation that are not available in the context we consider.  Our work demonstrates that traffic analyses attacks in the style of FATS are as effective for the current generation of consumer IoT devices as they were for sensor networks a decade ago. 

Our attack draws more broadly from the extensive literature on side-channel privacy attacks in the web and other contexts.  Several such attacks have been demonstrated on anonymity networks~\cite{back2001traffic, murdoch2005low}.  Internet browsing habits are also subject to privacy breaches by website fingerprinting attacks using network traffic metadata \cite{felten2000timing, gong2012website, wang2014effective}. Network traffic metadata has also been used to perform user and device fingerprinting by a variety of techniques, including correlating IP ID header fields~\cite{bellovin2002technique}, clock skews from the TCP timestamp option~\cite{kohno2005remote}, and hidden Markov models trained on Netflow data~\cite{verde2014no}. 

Our proposed traffic shaping solution was motivated by existing work on traffic shaping in anonymity networks. Independent link padding algorithms, which force traffic to match a predefined schedule or distribution, are well-studied~\cite{van2015vuvuzela, dyer2012peek}.  Fu, et al. developed a Bayesian derivation of detection rates for constant-interval-timing (CIT) versus variable-interval-timing (VIT) independent link padding~\cite{fu2003analytical}, and show that real-world system jitter makes CIT shaping ineffective for an adversary with lots of low-noise data. It is generally believed that independent link padding algorithms require too much latency or bandwidth overhead to be feasible in practice~\cite{dyer2012peek}. Our results challenge this notion in the IoT context. 

Many dependent link padding techniques that rely on feedback from shaped traffic have also been presented.  Wang et al. designed a dependent link padding algorithm that uses matched packet schedules to prevent an observer of a mix node in an anonymity network from pairing incoming flows with outgoing flows~\cite{wang2008dependent}.  Shmatikov et al. developed an adaptive padding algorithm that forces inter-packet intervals of short-lived web communications through an anonymity network node to match a pre-specified probability distribution~\cite{shmatikov2006timing}. Although such techniques require less overhead than independent link padding algorithms, most cannot prevent the attack we describe. Using feedback from shaped traffic, periods of higher or lower traffic rates are retained by dependent link padding algorithms.  This approach is acceptable in the anonymity network context where the goal is to prevent correlation of source and destination across multiple flows, but it is unacceptable if we wish to prevent identification of traffic correlated with user activities within a single flow. 

Previous work has also explored how sensitive properties of data can be hidden through statistical means. Researchers have used differential privacy, for instance, to place bounds on an attacker's ability to learn sensitive information through strategic database querying~\cite{dwork2008differential}. This inspires our own research: can we offer strong, provable guarantees of privacy from traffic analysis attacks on smart home devices?

\section{Threat Model}
\label{sec:threat}

Our analysis focuses on the abilities of a passive network observer with access to last-mile Internet traffic into and out of a smart home.  We consider our adversary to have capabilities similar to those of an ISP.  The adversary's goal is to infer users' in-home activities from smart home device network traffic. 

Traffic collection can occur at the flow level (using IPFIX or NetFlow) or at the individual packet level (storing data in a standard pcap format). 
ISPs regularly collect flow-level traffic records for network management purposes. 
The possibility of inferring sensitive user information from these records is especially worrisome as it would require no change to existing data collection procedures. 
Our proposed traffic shaping solution protects consumer privacy against adversaries with Netflow data or packet captures. 

Our adversary is not active and does not need to manipulate traffic to or from the smart home to conduct the attack. 
We assume that the adversary may or may not have access to local area network (LAN) traffic from within the smart home network.
Some ISPs, for instance, provide customers with a home gateway router owned by the ISP. 
An ISP with LAN data can more easily perform the traffic metadata privacy attack (Section~\ref{sec:attack}).
However, customers can purchase and connect their own router or Wi-Fi access point to the ISP-provided router, creating a new LAN invisible to the ISP. 
In either case, our proposed traffic-shaping solution can be implemented on a third-party hub/router or individual devices themselves.
Smart home device traffic is therefore shaped before reaching an adversary-controlled router.

We assume that packet contents are encrypted and the adversary must rely on traffic rate and packet header metadata to infer user activities. 
In fact, almost all IoT devices we tested use TLS/SSL when communicating with first and third party cloud servers. Given the increasing focus on security in the IoT community, encrypted communications will likely become standard for smart home devices. 
By ignoring packet contents, our privacy attack indicates that sensitive information about user activities is still at risk even when industry best practices for data encryption are in place.  

Finally, we assume that the passive network adversary can obtain a database of labeled traffic from smart home devices for training machine learning algorithms. We do not bound our adversary's ability to analyze example traffic and assume that they can continuously monitor a target smart home's traffic if it allows them to better conduct the traffic metadata attack.
\section{Experiment Setup}
\label{sec:setup}

We set up a laboratory smart home environment with several commercially-available IoT devices as a testbed for performing our traffic metadata attack and developing our privacy protection strategy. 

We configured a Raspberry Pi 3 Model B as an 802.11n wireless access point for use as the laboratory smart home's gateway router. 
The Raspberry Pi 3 has a built-in WiFi antenna, so no additional hardware was needed. 
The Raspberry Pi ran the Raspbian Jessie OS, a version of Debian Linux optimized for the Raspberry Pi platform.  
Hostapd (host access point daemon) enabled mac80211-compliant access point and authentication server services. 
Dnsmasq enabled DNS and DHCP services.  
An iptables network address translator (NAT) connected the Raspberry Pi's wireless interface to the wired interface connected by Ethernet to the Internet.  

This setup allowed us to record all network traffic to and from the smart home devices connected to the Raspberry Pi's Wi-Fi network as pcaps and IPFIX records.  The Raspberry Pi also provided a convenient platform for implementing our proposed traffic shaping solution.  Because many commodity home routers also run some version of Linux \cite{kofler2011evaluating}, this implementation is widely deployable.  

We included the following popular IoT devices in our laboratory smart home, covering a range of device types, manufacturers, and privacy concerns: 

\begin{enumerate}
\item Sense Sleep Monitor \cite{sense} 

\item Nest Cam Indoor security camera \cite{nestcam}.

\item Amcrest WiFi Security IP Camera \cite{amcrest}. 

\item Belkin WeMo switch \cite{wemo}. 

\item TP-Link WiFi Smart Plug \cite{tplink}.

\item Orvibo Smart WiFi Socket \cite{orvibo}.

\item Amazon Echo \cite{echo}. 
\end{enumerate}

The Sense Sleep Monitor infers users' sleep patterns and sleeping environment using built-in motion, light, temperature, air quality, and microphone sensors.
The Amcrest and Nest cameras are always-on video cameras that allow motion detection and live streaming via cloud-based web and mobile applications. 
The WeMo, TP-Link, and Orvibo devices are electrical outlets with an on/off switch that can be remotely controlled by a smartphone application. 
The Amazon Echo is a voice-controlled personal assistant that responds to user queries and can interface with other cloud-based services and IoT devices.

These devices are by no means exhaustive of the wide range of available IoT smart home products.  However, they encompass a variety of device types from large and small manufacturers. Given the effectiveness of traffic rate privacy attacks on all tested devices, we believe that smart home owners should be concerned about traffic rate metadata across all types of smart home devices.  
\section{Traffic Rate Privacy Attack}
\label{sec:attack}

The smart home traffic rate metadata attack has two components: device identification and activity inference. 

At a high level, the adversary leverages devices' known specific purposes to map changes in traffic rates to user activities.  As a simple example, an adversary might first identify that a particular traffic flow is from a smart outlet.  Smart outlets generally  have only two functions (turn on and turn off), so if the flow indicates a spike in traffic rate at a particular time, the adversary can infer that the outlet was turned on or off at that time. The following subsections describe each step in this attack in detail with case studies from commercially-available smart home devices. 

\subsection{Device Identification}
\label{sec:device-identification}
First, the adversary associates individual traffic flows with a known type of device. We mean "traffic flow" in the same sense as it is used in NetFlow: a flow is a set of packets associated with a 5-tuple of \texttt{sender\_ip}, \texttt{recipient\_ip}, \texttt{sender\_port}, \texttt{recipient\_port}, and \texttt{protocol} within some time window. There are several methods to perform device identification depending on the information available to the attacker.  

\subsubsection{Using MAC Addresses} 
An attacker with smart home LAN data can use the first three bytes of device MAC addresses (the organizational unique identifier) to assign manufacturer labels to each flow. 
Knowing devices' manufacturers makes the following DNS and traffic rate techniques easier because the space of possible device identities is reduced. 
However, the following techniques are still effective for a last-mile observer without device MAC addresses. 

\subsubsection{Using DNS Queries} 
DNS queries associated with each flow can often be associated with a particular device.  We recorded DNS queries from our smart home laboratory and found that 4 out of 7 devices could be uniquely identified by inspection of DNS queries alone: the Sense Sleep Monitor, the Nest Camera, the Amcrest Security Camera, and the Amazon Echo (Figure~\ref{fig:dns-queries}).  
For example, the Nest Camera queried domains from dropcam.com (the predecessor to the Nest Camera), while the Sense sleep monitor queried domains from hello.is (the company that makes the Sense).
An adversary could use a laboratory setup like our own to learn these mappings or perform reverse DNS lookups to pair service IPs with device-identifying domain names.

\begin{figure}[t]
\begin{center}
\small
\begin{tabular}{ll}
\hline
\textbf{Device} & \textbf{DNS Queries} \\
\hline
Sense Sleep Monitor & \url{hello-audio.s3.amazonaws.com} \\
                                  & \url{hello-firmware.s3.amazonaws.com} \\
                                  & \url{messeji.hello.is} \\
                                  & \url{ntp.hello.is} \\
                                  & \url{sense-in.hello.is} \\
                                  & \url{time.hello.is} \\
\hline
Nest Security Camera & \url{nexus.dropcam.com} \\
                                    & \url{oculus519-vir.dropcam.com} \\
\hline
Amcrest Security Camera & \url{amcrestcloud.com} \\
                                &
                         \url{command-3.amcrestcloud.com} \\
                                & \url{ftp.amcrestcloud.com} \\
                                & \url{media-amc-1.hostedcloudvideo.com} \\
                                & \url{p2p.amcrestview.com} \\
                                & \url{dh.amcrestsecurity.com} \\
\hline
Belkin WeMo Switch & \url{prod1-fs-xbcs-net-1101221371.}\\
                       &\quad \url{us-east-1.elb.amazonaws.com} \\
                       & \url{prod1-api-xbcs-net-889336557.} \\
                       &\quad \url{us-east-1.elb.amazonaws.com} \\
\hline
TP-Link Smart Plug & \url{devs.tplinkcloud.com} \\
                        &
                    \url{uk.pool.ntp.org} \\
\hline
Orvibo Smart Socket & \url{wiwo.orvibo.com} \\
\hline
Amazon Echo & \url{ash2-accesspoint-a92.ap.spotify.com} \\
                       & \url{device-metrics-us.amazon.com} \\
                       & \url{ntp.amazon.com} \\
                       & \url{pindorama.amazon.com} \\
                       & \url{softwareupdates.amazon.com} \\
\hline
\end{tabular}
\caption{DNS queries from smart home devices during a representative packet capture.  Many queries can be easily mapped to a specific device or manufacturer.}
\label{fig:dns-queries}
\end{center}
\end{figure}

However, the remaining 3 devices issued DNS queries for domains that could only be associated with a manufacturer (but not a particular device) or with a third-party cloud hosting service (e.g. Amazon AWS).  Flows from these devices are more difficult to identify, but we found that all devices in this category issued DNS queries for more than one domain.  Therefore, the set of domains may be unique for each device from a particular manufacturer, providing a device-type fingerprint that can be used for identification. An adversary could learn the unique arrangement of domains used by each device and use them to identify flows in NetFlow records or packet captures. 

\subsubsection{Using Traffic Rates}
\begin{figure}[t]
\centering
\includegraphics [width=0.42\textwidth]{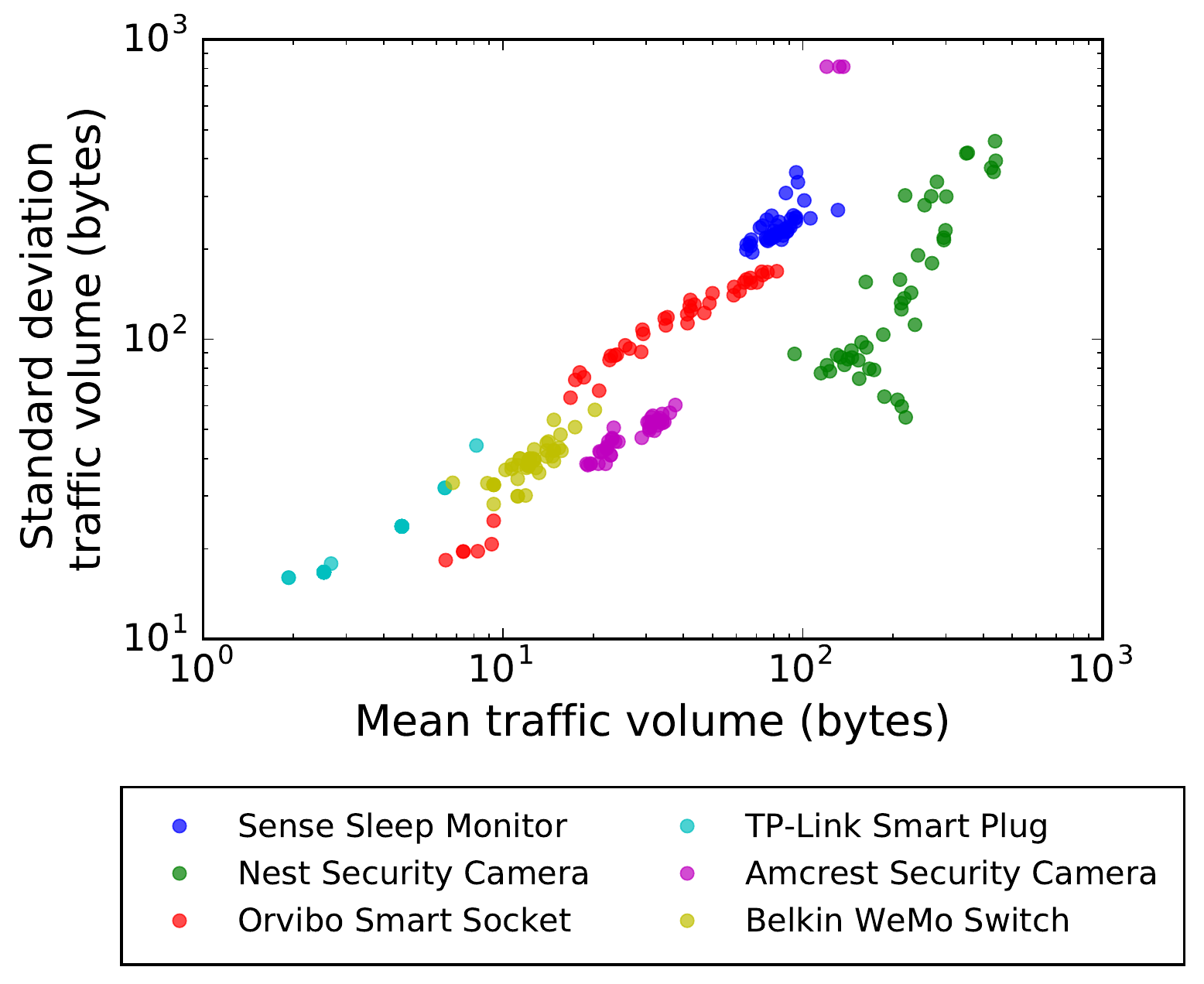}
\caption{Internet traffic rates to and from individual smart home devices clusters well by the mean and standard deviation of traffic volume within a several-minute time window.}
\label{fig:clustering}
\end{figure}

\begin{figure}[t]
\centering
\includegraphics[width=0.42\textwidth]{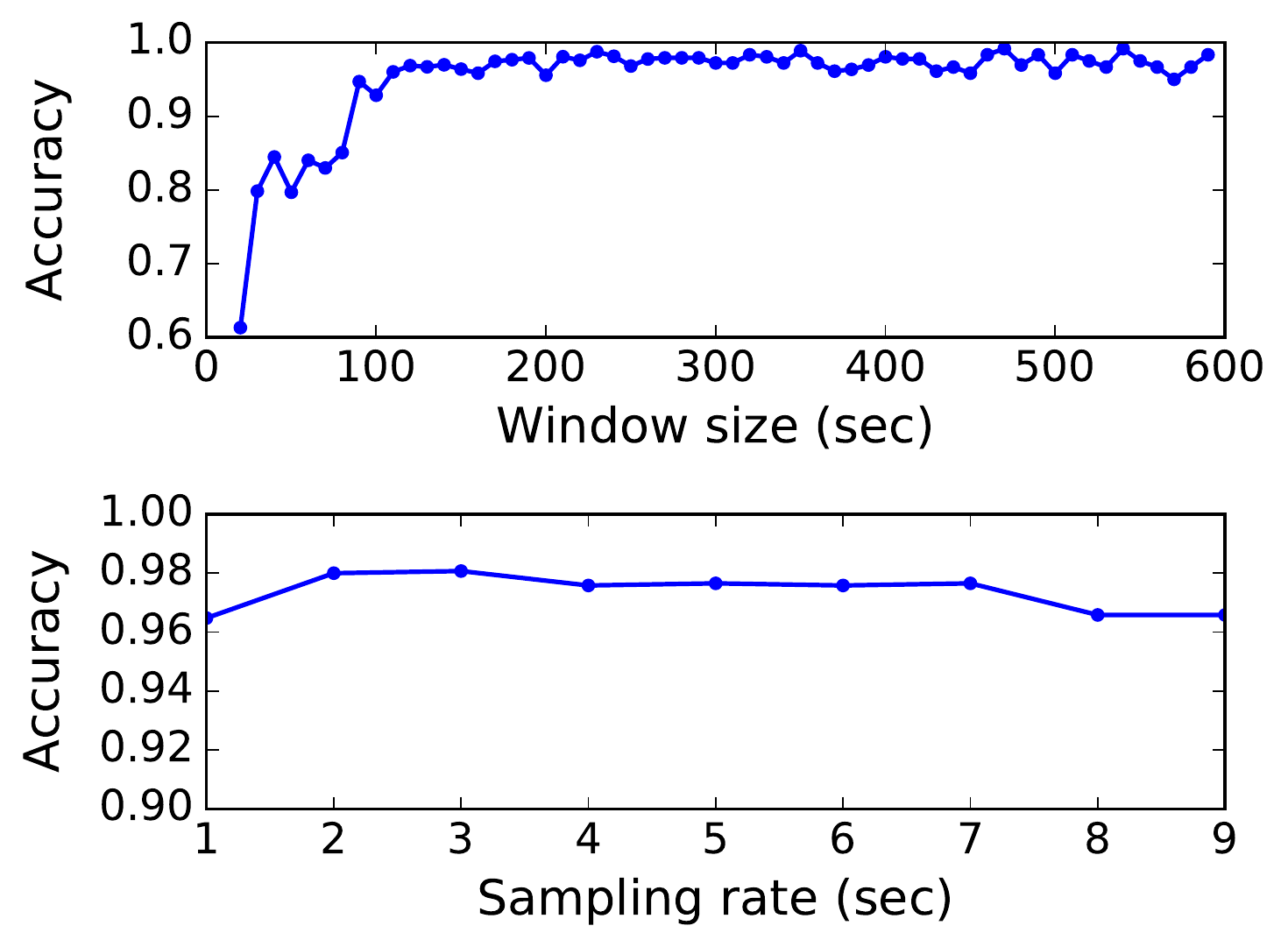}
\caption{
Supervised machine learning can accurately fingerprint smart home devices using Internet traffic rates. 
These plots show the accuracy of a $3$-nearest-neighbors classifier identifying which of six commercially-available smart home devices generated the traffic within a time window for varying window size and sampling rate.
Accuracies determined by 10-fold stratified cross-validation.}
\label{fig:classification}
\end{figure}

If DNS queries and/or MAC addresses are insufficient for identifying the devices that correspond to traffic flows, 
it is likely still possible to fingerprint devices using traffic rates.
As a proof of concept, we collected two hours worth of network traffic from six smart home devices operating with minimal user interactions.  
We separated this traffic into flows corresponding to individual devices and converted into time series vectors. Vector elements contained the amount of traffic (bytes) sent or received by the device in consecutive \mbox{$s$-second} samples. 
We further divided these vectors into \mbox{$w$-second} windows and 
extracted \mbox{2-element} feature vectors containing the mean and standard deviation of the traffic amounts in the samples of each window.  
These feature vectors clustered well by the device responsible for the traffic (Figure~\ref{fig:clustering}). 
A $3$-nearest-neighbors classifier trained on these feature vectors resulted in greater than $95\%$ accuracy classifications for a range  of $w$ and $s$ (Figure~\ref{fig:classification}). 
Accuracies were determined by 10-fold stratified cross-validation on the training set. 

Surprisingly, simple traffic features can distinguish devices that belong to the same category of smart home appliance.
For example, the mean traffic volumes of the Nest and Amcrest security cameras differed by almost an order of magnitude over the data collection period. 

In more complex scenarios there are many additional features of network traffic that could be used for device identification \cite{karimi2016distributed}. These include the proportion of SYN and ACK packets per flow, the number of packets in a flow, and distributions of inter-packet intervals. 
There are also a variety of machine learning techniques beyond the k-nearest-neighbors algorithm we tested that could be applied to device identification.  
An adversary can leverage any number of features or algorithms depending on the smart home devices they wish to identify and prior knowledge of their target. 

Future research using data from an ISP could develop more advanced techniques for smart home device fingerprinting on a larger scale.

\subsection{Activity Inference from Traffic Rates}
We performed a series of controlled experiments to test the network behavior of smart home devices in response to specific privacy-sensitive user activities. 
For all of the tested devices, changes in traffic rate correlated to device state changes caused by user activities.
An adversary can use this knowledge after device identification to infer user activities from changes in traffic rates. 

\begin{figure}[!t]
\centering
\includegraphics[width=.49\textwidth]{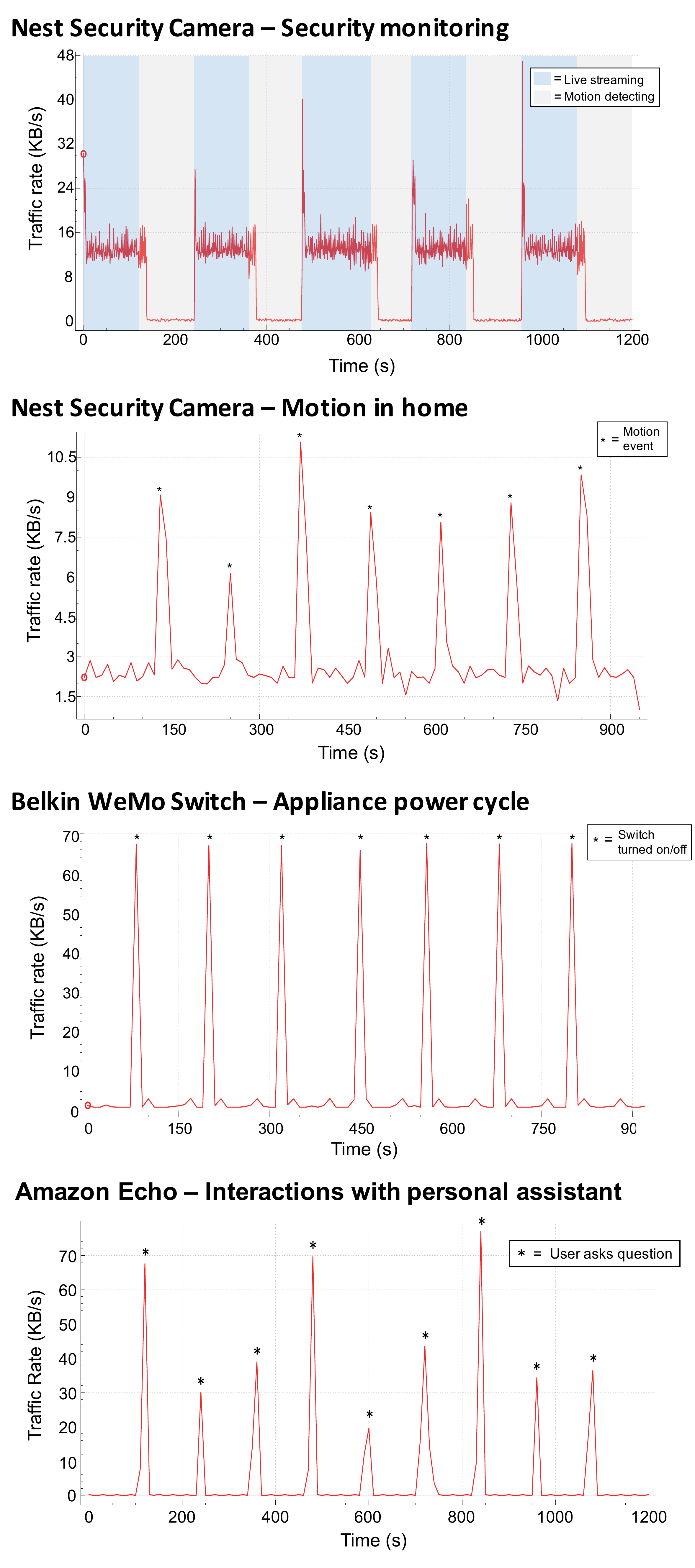}
\caption{Network traffic send/receive rates of selected flows from 4 commercially-available smart home devices during controlled experiments. Clearly visible changes in send/receive rates directly correspond with user activities.  A passive network observer aware of this behavior could easily correlate smart home traffic rates with device states and corresponding user activities. \vspace{12pt}}
\label{fig:pcap-attack}
\end{figure}

\textbf{Belkin WeMo Switch.} The WeMo switch only has two states, on and off, and the network send/receive rates reflect this binarity. Figure~\ref{fig:pcap-attack} shows WeMo network behavior when the switch is turned alternatively on and off every 2 minutes using the WeMo smartphone app and the physical button on the device. The spike in traffic every time the switch changes state clearly reveals user interactions with the device to an network observer.

\textbf{Amazon Echo.} We tested the Echo by asking a a series of 3 questions (``what is the weather?,'' ``what time is it?,'' and ``what is the distance to Paris?'') repeated 3 times, one question every 2 minutes. Figure~\ref{fig:pcap-attack} shows the send/receive rates of SSL traffic between the Echo and a single \texttt{amazon.com} IP address during the experiment. Although the Echo sent and received other TCP traffic to different domains during this time, we were able to identify the stream that correlated with the questions. A network-level attacker with their own test device can also identify this stream and use the SSL traffic spikes to infer when user interactions occurred.

\textbf{Sense Sleep Monitor.} Figure~\ref{fig:sleep-pcap} shows send/receive rates from the Sense over a 12 hour period from 10:40pm to 10:40am. Notably, the send/receive rate peaked at times corresponding with user activity. The user shut off the light in the laboratory smart home and went to bed at 12:30am, temporarily got out of bed at 6:30am, and got out of bed in the morning at 9:15am. 

We believe that the ability of an adversary to tell when a user is sleeping, or at least in bed, from network send/receive rates constitutes a significant privacy vulnerability.
ISPs can already guess that users are sleeping when network traffic from smartphones or PC web browsers decreases at night; however, this relies on many assumptions, e.g. that users only stop using their other devices immediately prior to sleeping, that everyone in the home sleeps at the same time and does not share other devices, and that users do not leave their other devices running to perform network-intensive tasks or updates while they sleep.
The single-purpose IoT nature of the Sense sleep monitor makes none of these assumptions necessary to infer users’ sleeping patterns from Sense traffic.

\textbf{Nest Security Camera.} Our observations of the Nest camera indicate that it has at least two primary modes of operation: a live streaming mode and a motion detection mode. In the live streaming mode, the camera’s video feed is either being actively viewed by the user through the Nest web/mobile application or the feed is being uploaded in real time to be stored on the cloud. In the motion detection mode, the video stream is not being uploaded, but the camera is monitoring the stream locally for movement. If movement is observed, the camera records a snapshot of the video and alerts the user.

Figure~\ref{fig:pcap-attack} shows send/receive rates from the Nest camera alternating between live streaming and motion detection mode every 2 minutes. The traffic rate is orders of magnitude higher in live streaming mode (and a short time afterward until the camera is notified that the user has stopped viewing the stream), allowing an adversary to easily determine whether or not the camera’s live feed is being actively viewed or recorded. 

Figure~\ref{fig:pcap-attack} also shows that an adversary could easily determine when a Nest camera detects movement when it is in motion detection mode. The camera was pointed at a white screen with a black square that changed location every two minutes. These simulated motion events triggered clearly observable spikes in network traffic. This predictable variability in network send/receive rates would allow an adversary to observe the presence and frequency of motion inside a smart home.

These issues are significant privacy vulnerabilities and physical security risks even though the content of the video stream remains protected by encryption. It should not be possible for a third party to be able to determine when a security camera detects movement or is being actively monitored.

\section{Evaluating Current Defenses}
\label{sec:solution}

In this section, we evaluate several existing techniques known to prevent traffic analysis attacks in other contexts.

\subsection{Blocking Traffic}
One strategy for preventing activity inference is to prevent an adversary from collecting smart home network traffic in the first place. 
The most naive method is to deploy a firewall that prevents all IoT device traffic from leaving a smart home's local network. 
Configuring such a firewall is straightforward on an intermediate hub-like device or the home gateway router.
This approach, however, rests on an untested assumption: that devices do not require cloud services to meet some basic functionality.

We tested the effect of removing Internet connectivity from seven commercially-available devices  while maintaining the local network (Figure~\ref{fig:blocking}). 
Four of the devices retained limited functionality, although without many of the ``smart'' features which distinguish these devices from their non-IoT analogs.
The remaining three devices had no functionality without Internet connection and were completely unusable.
This was surprising, because there is no technical reason why these devices could not have continued to partially function.
The sleep monitor still could have reported current sleeping conditions detected using on-device sensors. 
The security cameras could have allowed a smartphone also on the local network to view the video feed (e.g., for monitoring an infant while in another room). 

\begin{figure*}
\begin{center}
\small
\begin{tabular}{l|l|l}
\textbf{Device} & \textbf{Functionality} & \textbf{Description} \\
\hline
Amazon Echo & limited & Can use as a bluetooth speaker with previously paired smartphone \\
&& Echo recognizes ``Alexa" keyword but does not provide any voice-control features \\ \hline
Belkin WeMo Switch & limited & Can turn switch on/off with physical button on device \\ 
&& Cannot use smartphone app to control device even when phone on local network\\ \hline
Orvibo Smart Socket & limited & Can turn switch on/off with physical button on device or smartphone app on local network \\ \hline
TP-Link Smart Plug & limited & Can turn switch on/off with physical button on device or smartphone app on local network \\ \hline
Nest Security Camera & none & Unable to view video feed or receive detected motion notifications \\ \hline
Amcrest Security Camera & none & Unable to view video feed or control camera direction\\ \hline
Sense Sleep Monitor & none & Monitor does not record sleep data\\&& Light-based UI does not reflect local sensor readings \\
&& Cannot use smartphone app to control device or access current data \\
\end{tabular} 
\caption{All tested commercially-available IoT devices had limited or no functionality when firewalled to prevent communication outside of the smart home LAN. This suggests that device manufacturers should be encouraged to improve their ``minimum reliable product.'' }
\label{fig:blocking}
\end{center}
\end{figure*}

Rather than blocking all traffic indiscriminately, a hub or router could use more advanced analysis to discern what specific traffic from each device can be blocked safely.
If done correctly, smart home devices could function effectively while particularly sensitive traffic streams are prevented from leaving the home.
One possible approach is for device developers to intentionally separate information sent to cloud servers into multiple streams categorized by information type and/or necessity for device functionality. 
A firewall could then block specific streams to trade-off users' privacy versus usability preferences. 

However, developers are likely disincentivized to allow users to block specific information streams even if if is technically feasible for them to do so. 
Without developer support, a third-party hub or router could potentially learn what packets or flows from particular devices can be blocked without excessively limiting functionality. 
However, encrypted traffic contents and the potential for delayed device failures makes this an open research question. 

\subsection{Tunneling traffic}
\label{sec:tunneling}
Another strategy for preventing device identification and activity inference is to tunnel all smart home traffic through a virtual private network (VPN).
This would impede a last-mile observer's ability to split traffic into individual flows and associate those flows with specific devices.

A VPN wraps all traffic from an endpoint in an additional transport layer, aggregating it into a single flow.
This flow has the source IP address of the home gateway router and the destination IP address of the server implementing the VPN exit point.
In effect, an adversary would see all traffic as originating and terminating from a single pair of endpoints, rather than from individual smart home devices (or the public IP address of the smart home NAT) and individual cloud servers.
That aggregation could make device identification more difficult.
When many IoT devices send traffic simultaneously, noise variations in traffic rates of high-rate devices could mask the entire traffic patterns of other devices. 
VPNs in general make it difficult to determine which variations in the overall traffic rate observed from outside a VPN correspond to individual devices, if the adversary can even identify what those individual devices were.

However, this increased attack difficulty does not give a consumer any actual proof of privacy protection. 
We have identified three cases where an adversary can infer activity despite seeing only VPN traffic. In each of these cases, an adversary can still fingerprint devices using the rate of VPN traffic:

\textbf{1. Single device.} 
If a smart home has only one IoT device, the VPN traffic rate will match the traffic from that device. Although the adversary cannot use DNS for device identification, they can leverage other traffic rate signals for device fingerprinting. The attack can then proceed as before.

\textbf{2. Sparse activity.} 
If there are multiple devices that send traffic at different times, an adversary could still identify events corresponding to single devices, since many time windows would contain traffic from only a single device. 
For example, a smart door lock and smart sleep monitor are less likely to be recording user activities simultaneously. 
Traffic observations from particular times of day are likely to contain non-background traffic from only one of these devices.
This will allow an adversary to identify the active device within a time window and perform activity inference as before. 

\textbf{3. Dominating device.} Even if device traffic overlaps, an observer could infer the presence of the device that sends the most traffic if it significantly overshadows traffic from other devices. 
For example, traffic from a security camera uploading live video will dominate any traffic from less network-intensive devices such as smart outlets. 
The adversary will be able to perform the attack on the high traffic rate device unaffected by the relatively small amount of traffic from the other devices. 

At minimum, these three cases demonstrate that a VPN can still reveal  traffic rate patterns that make the metadata privacy attack possible.
This motivates more technically sophisticated solutions that prevent the leakage of any information about user activities.
\section{Traffic Shaping}
\label{sec:shaping}

A user cannot reasonably block smart home traffic from leaving their home without making their devices unusable.
While tunneling via a VPN does make the attack somewhat more difficult, it does not provide any guarantee of privacy; a dedicated adversary could still use the available traffic rate information from the VPN in machine learning.
Traffic shaping is the only solution that can prevent the leak of rate information and thus render the attack impossible.
We propose traffic shaping using independent link padding (ILP). 
ILP involves shaping traffic rates to match a predetermined rate or schedule, thereby exposing no information about device behavior to an adversary.
For a relatively low-cost, traffic shaping can give a consumer strong guarantees of smart home privacy without interfering with the basic functioning of smart home devices.

\subsection{Overview}

\begin{figure}[t]
\centering
\includegraphics[width=0.43\textwidth]{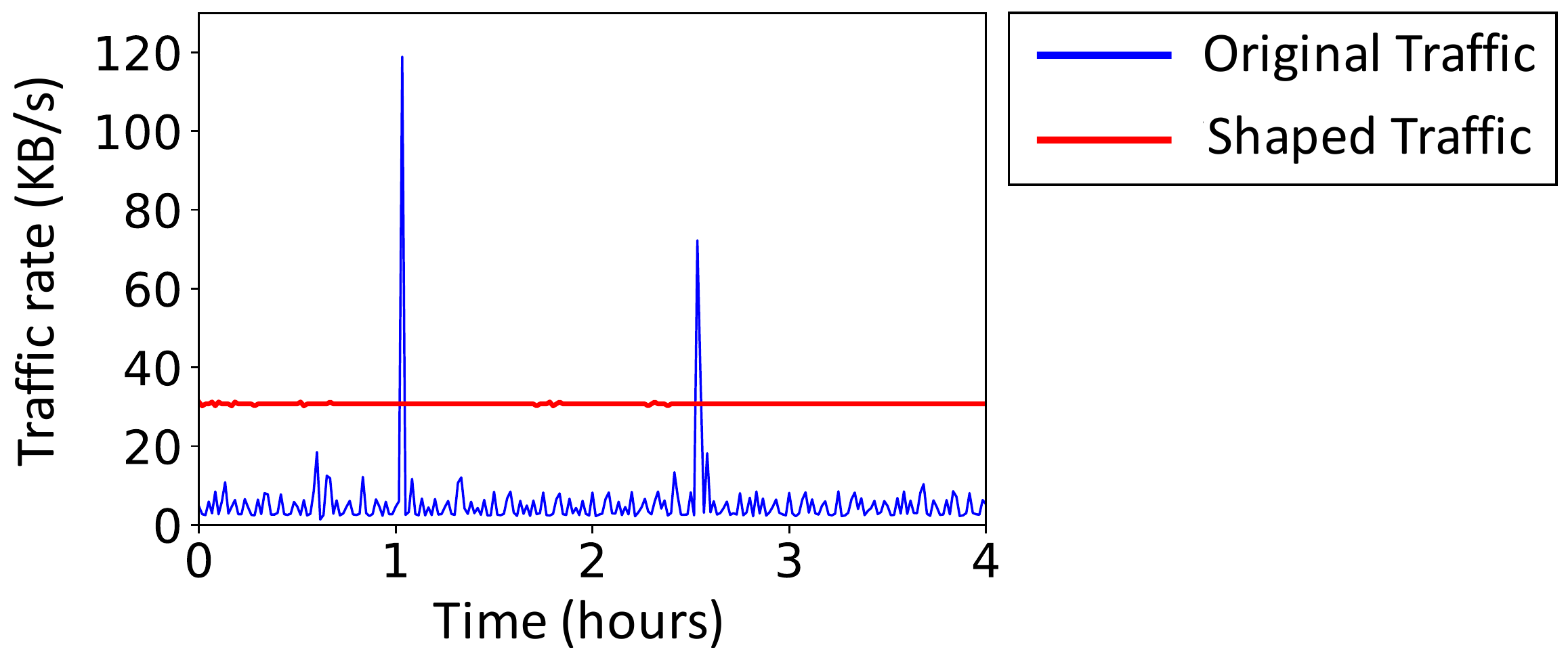}
\caption{ Results of simulated traffic shaping by independent link padding on combined traffic from 3 smart home deices: a Belkin WeMo Switch, an Amazon Echo, and a Sense Sleep Monitor. All packets were fragmented or padded to 512 bytes. Inter-packet intervals were drawn from a normal distribution with mean 1s and standard deviation 10ms.  Note that traffic shaping masks traffic spikes which correspond to user activities, preventing the described traffic metadata privacy attack.}
\label{fig:shaping-example}
\end{figure}

One approach to traffic shaping that can guarantee smart home privacy involves sending fixed-size packets at a constant rate independent of the underlying device traffic.  This is a form of ILP that results in traffic rate metadata with no information about device states and corresponding user activities (Figure~\ref{fig:shaping-example}). 

If the shaped send rate is lower than the rate of device traffic, packets from the devices are delayed in a queue until they can be sent at the shaped rate.
If the shaped rate is higher than the rate of device traffic, fake packets (cover traffic) are added.    

This results in a trade-off between latency and bandwidth overhead.  
If a high shaped traffic rate is chosen, little additional latency will be imposed on device traffic, but lots of bandwidth will be needed for cover traffic.
If a low shaped traffic rate is chosen, less cover traffic is needed, but device traffic could experience long latencies.  The shaped traffic rate should therefore be chosen to minimize the amount of cover traffic needed while imposing latency that doesn't interfere with device function.

\subsection{Traffic Shaping Implementation}
To evaluate the device latency tolerances and corresponding traffic shaping bandwidth costs, we built a working traffic shaping implementation that we deployed on our Raspberry Pi WiFi access point in our smart home laboratory (Section~\ref{sec:setup}). 
The implementation consists of two components: a VPN tunnel and a traffic shaper.

\textbf{VPN tunnel.} 
We use OpenVPN on an Amazon EC2 \texttt{t2.medium} instance as the endpoint for traffic originating from our smart home access point. Though a VPN alone does not provide any guarantee of privacy against the traffic rate metadata attack, it is a necessary component of our implementation. Traffic shaping requires that we send cover traffic when no real packets are available. 
The VPN renders the real traffic packet headers indistinguishable from cover traffic packet headers. 

The VPN also allows us to shape traffic going to the smart home (download traffic) from Internet services.
We run the same traffic shaping implementation on the VPN endpoint to prevent an adversary from inferring user activities from download traffic metadata. 

The only other requirement of the VPN is that it sit outside of the last-mile observer's view; for an adversary like an ISP, this requires that the VPN have a different service provider than the smart home itself.

\textbf{Traffic shaper.} We built our traffic shaping implementation around the Linux kernel's traffic control systems, configurable via the \texttt{tc} command line tool. Through traffic control, a user can modify the kernel's strategy for how it queues and sends network packets. 
We use it to enforce maximum transmission rates and prioritize device traffic over cover traffic on the network interface responsible for the VPN tunnel, \texttt{tun0}. 

\begin{figure}[t]
\centering
\includegraphics[width=0.45\textwidth]{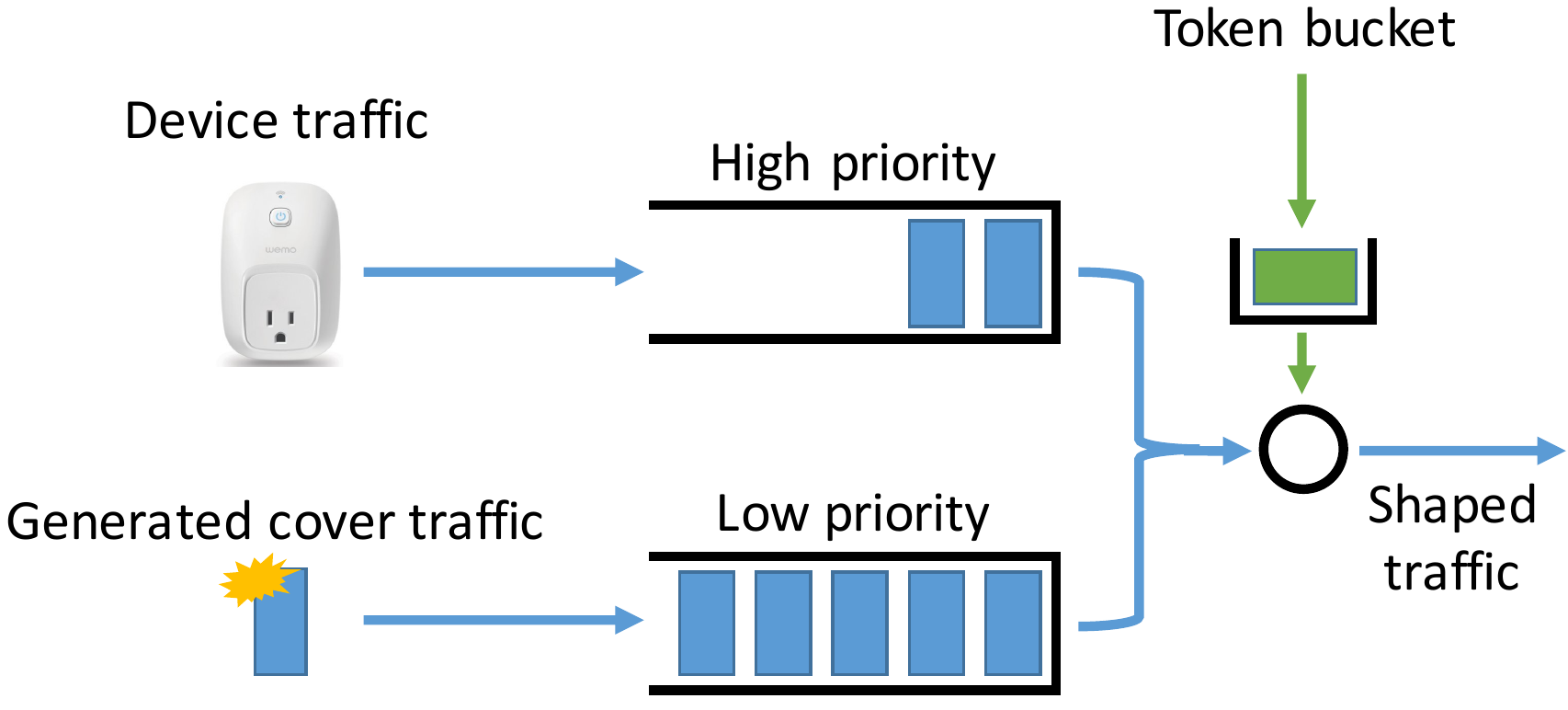}
\caption{The components of our traffic shaping implementation, including cover traffic generation and traffic control in the kernel. The token bucket shaper has a buffer size of 1 token, making it equivalent to a fixed-rate leaky bucket. The rate of token arrival is set to the desired shaped traffic rate. Device packets in the high priority queue are always sent before cover packets in the low priority queue.}
\label{fig:shaper_diagram}
\end{figure}

Figure~\ref{fig:shaper_diagram} shows how the traffic shaper is configured to shape device and cover traffic so that it is sent through the network interface at the required rate. 

First, a user space program generates UDP cover traffic sent to a dummy IP address. 
The VPN endpoint drops all cover traffic sent to this address at the end of the tunnel. 

Next, we create a new PRIO queuing discipline (qdisc). The PRIO qdisc has three priority classes for packets. If packets are available in a high-priority queue, the qdisc will send those packets before sending packets from a lower-priority queue. We set a traffic control filter so that device traffic is assigned to a high priority queue, and generated cover traffic is sent to a lower priority queue. 
Therefore, device traffic always takes precedence over cover traffic.
Cover packets are generated at a faster rate than the shaped traffic rate so there is always a cover packet ready to be sent if a device packet isn't available. 

Finally, we replace the default first-in-first-out queuing discipline of \texttt{tun0} with a hierarchical token bucket (HTB). We set the HTB as the parent of the PRIO qdisc. When a packet is ready to transmit from the PRIO qdisc, either from the high or low priority queue, the interface attempts to grab a token from the HTB. If a token is available, the packet is sent. Otherwise, the interface must wait for a token to become available.  
By setting the rate at which the HTB generates tokens and the maximum number of tokens available, we can enforce a fixed maximum rate of traffic on the interface\footnote{Our configuration of the hierarchical token bucket effectively makes it a fixed-rate leaky bucket.}.

Together, these three steps ensure that:
\begin{enumerate}
\item Traffic originating from the hub is sent out at a constant rate, regardless of the activity happening within the LAN 
\item Device traffic takes precedence over cover traffic to minimize device traffic latency.
\end{enumerate}

Our script to perform cover traffic generation and configure the kernel's traffic control takes less than 100 lines of code. While we deployed this on our own Raspberry Pi hub, it is readily-portable to any hub that uses a Linux-based kernel. 

\subsection{Devices Tolerate High Latencies}
We tested seven devices with our traffic shaping implementation to determine how much shaping-induced latency the devices could tolerate. 
We varied the shaped traffic rate, measuring the average latency experienced by all device packets and the maximum latency experienced by any individual device packet. We were able to find the lowest rate (and corresponding latency overhead) at which each device still functions. 
The least latency-tolerant device, the Amazon Echo, tolerated 0.6 seconds of average per-packet latency and 1.4 seconds of maximum latency. The most tolerant device, the Sense Sleep Monitor, tolerated 6.2 seconds of average per-packet latency and 14.4 seconds of maximum latency.
Figure~\ref{fig:implementation-results} reports latency measurements for all tested devices at their lowest tolerated shaped traffic rate and the effect of slower traffic on device function. 
All tolerances are considerably longer than the $<$100ms of typical per-packet network latency without traffic shaping. 

\begin{figure*}[!t]
\small
\begin{tabular}{$L{3cm}^L{1.5cm}^L{1.5cm}^L{1.3cm}^L{1.3cm}^L{6.3cm}} 
\textbf{Device} & \textbf{Shaped upload traffic rate} & \textbf{Shaped download traffic rate} & \textbf{Avg. per-packet latency } & \textbf{Max. per-packet latency} & \textbf{Effects of traffic shaping} \\
\hline
Amazon Echo & 10 KB/s & 10 KB/s & 0.9s & 4.4s & Echo answered questions with several seconds delay. With slower traffic, the Echo does not completely answer the question.   \\ \hline
Belkin WeMo Switch & 5.0 KB/s & 2.5 KB/s &  1.3s & 2.6s & Works with few seconds of delay when switching from smartphone app. With slower traffic, the app loses connection to the switch. \\ \hline
Orvibo Smart Socket & 0.25 KB/s & 0.25 KB/s & 0.8s & 2.6s  & Works with few seconds of delay when switching from smartphone app. With slower traffic, the app registers switching success, but the socket does not actually change state. \\ \hline
TP-Link Smart Plug & 0.5 KB/s & 0.5 KB/s & 0.8s & 1.7s & Works with few seconds of delay when switching from smartphone app. With slower traffic, the app reported that the plug was unreachable.  \\ \hline
Nest Security Camera & 10 KB/s & 10 KB/s & 8.5s & 17.6s  & Video streamed with 10-15 seconds of lag and app reported slow upload speeds. With slower traffic the app lost connection with the camera. \\ \hline
Amcrest Security Camera & 35 KB/s & 2.5 KB/s & 1.5s & 3.4s & Video streamed with 1 minute delay, but no breakage. Could tolerate slower traffic, but with >3 minute delay and less reliability. \\ \hline
Sense Sleep Monitor & 2.5 KB/s & 2.5 KB/s & 0.4s & 1.4s & The smartphone app could still receive live updates. With slower traffic, live updates stop.\\ 
\end{tabular} 
\caption{Lowest tested shaped traffic rates and corresponding per-packet latencies at which smart home devices still function, dependent on usage patterns and traffic load on the shaper.
Reported per-packet latencies were measured in the upload direction.  Most tested device were more latency-tolerant for shaping in the download direction.
If a device is idle for long periods of time, the amount of cover traffic overhead added on top of device traffic can equal the overall shaped traffic rate.
This would result in the most bandwidth overhead for traffic shaping, because no user activities are being concealed.
}
\label{fig:implementation-results}
\end{figure*}

\subsection{Shaping Incurs Acceptable Throughput Overhead}
The bandwidth overhead of traffic shaping depends on the shaped traffic rate and the network behavior of individual devices.
We measured the cover traffic bandwidth required to shape at rates resulting in the highest network latency tolerated by each device. If a device is idle for long periods of time, the amount of cover traffic overhead added on top of device traffic can equal the overall shaped traffic rate (Figure~\ref{fig:implementation-results}).
This would result in the most bandwidth overhead for traffic shaping, because no user activities are being concealed. 
Unsurprisingly, devices of similar type required similar shaped traffic rates.

\textbf{Smart outlets/switches.} Both the Orvibo Smart Socket and TP-Link Smart Plug were very tolerant of low traffic rates. The Orvibo withstood traffic shaping to a rate of 0.25KB/s for upload and 0.25KB/s download, while the TP-Link only required 0.5KB/s for upload and 0.5KB/s for download. Traffic shaping at the rate required by the TP-Link switch would only consume about 2.5GB of extra data per month. This suggests that similarly low-bandwidth devices that do not upload audio or video data could be successfully shaped at a very low cost.  

\textbf{IP cameras.} Shaping smart home traffic that includes an IP camera is understandably more costly due to the bandwidth requirements of streaming video. The Nest security camera required a shaped bandwidth of at least 10KB/s upload and 10KB/s download in order to maintain its functionality. The Amcrest camera, on the other hand, experienced a more graceful degradation when faced with slower traffic rates. We did not find a single point at which it was completely ``broken.'' Rather, the video feed gradually got more and more laggy. A user could decide to trade-off more video lag for a lower shaping data overhead.

\textbf{Personal assistant.} The Amazon Echo also experienced a graceful degradation---as we lowered the shaped traffic rate, the Echo would experience several seconds of lag answering questions but was still functional. However, when the traffic rate was shaped to below 10KB/s upload and download, audio responses to user questions would stop mid-playback. 

\subsection{Shaping is Feasible for Many Users}
These findings demonstrate -- perhaps counterintuitively -- that traffic shaping can be a cost-effective means to guaranteeing privacy. 

\textbf{Network performance cost.} According to the Speedtest Market Report \cite{speedtest}, the average upload speed for fixed broadband in the United States was 2.36MB/s (18.88Mb/s) as of June, 2016. For users with at least 2MB/s upload speed, a bandwidth overhead  $\le 40$KB/s (above the required shaping rate for any of our tested devices) amounts to $\le 2\%$ of the user's upload capacity. This is low enough for shaping to protect all IoT devices in a smart home from traffic metadata privacy attacks without noticeably affecting Internet performance of non-IoT traffic (e.g., traffic from a personal computer or smartphone). 

Unfortunately, there are certain regions where traffic shaping may still be infeasible. Reports often cite India as having some of the slowest connection speeds in the Asia-Pacific region, with 1.7 megabits per second average download speed in 2014 \cite{india-internet-speed}. Traffic shaping under those network conditions would consume a significant portion of a consumer's entire bandwidth budget.

\textbf{Data usage cost.} Consumers with data caps may not want to risk hitting a data cap and paying a premium for extra data. Here, the cost of privacy can be put into real monetary terms: what is the financial cost of the data necessary to do traffic shaping? Our tests show that the data usage costs of traffic shaping are well within reach for consumers. 

Many smart home devices have network behavior similar to smart outlets and sleep monitors. For these classes of devices that do not stream audio or video, the data usage cost of shaping is very low.
Traffic shaping that adds 7.5KB/s bandwidth overhead, enough to mask non-audio/video devices like smart outlets, would only amount to 19GB of extra data used per month.
More data-intensive devices like cameras or baby monitors that stream audio or video require a greater amount of a data cap.
Traffic shaping that adds 40KB/s bandwidth overhead, enough to mask these data-intensive devices with a high level of performance, would count towards 104GB of a data cap in a month. This is a worst-case scenario: in our experiments, we did not find a device that required nearly that much bandwidth overhead.

Though 104GB in a worst-case scenario sounds substantial, it is feasible within many home Internet data plans around the world.
Comcast, the largest ISP in the United States by number of subscribers, imposes a one terabyte per month limit on consumer home Internet usage in at least 34 US states \cite{comcast-data-cap}. 
Above that, data costs \$10 per 50GB. 
Traffic shaping that adds 104GB of data would only count towards 10\% of a 1TB data cap in a month.

A consumer can even use traffic shaping in countries with more stringent data caps.
Internet service provider Telstra in Australia and Bell Internet in Canada, largest in their respective countries, offer affordable plans to users that have sufficiently large or unlimited Internet usage limits \cite{telstra-plans, bell-internet}.
The mid-tier plan offered by Bell Internet, for instance, provides 350GB of monthly usage, 25MB/s download speeds, and 10MB/s upload speeds \cite{bell-internet}.
A consumer with a 350GB data cap could traffic shape non-data-intensive device traffic with only 4\% of their overall data cap. If they wish to shape a camera or similarly data-intensive device they would have to consider the tradeoff between their desired level of privacy and data usage carefully. Ultimately, privacy through traffic shaping is still achievable with their data cap.

However, there are still many contexts in which consumers may not be able to afford the required overhead of ILP traffic shaping. 
Future research could explore alternative methods of traffic  shaping that trade-off privacy for bandwidth usage.  Such techniques might reveal coarse-grained user information (such as when a user is at home), but hide high-resolution details of user activities. 

\subsection{Real-world Deployment Considerations}

\textbf{Constant versus variable rate shaping}
The ILP traffic shaping implementation we use to evaluate latency and bandwidth costs is based on an algorithm that shapes all outgoing traffic to a constant rate.
In application, constant rate traffic shaping provides inadequate protection against some types of adversaries because it is impossible to actually send traffic at a perfectly constant rate. 
Due to ``system jitter,'' the shaped traffic will have very small variations in rate that distinguish periods of high or low cover traffic. 
This allows an adversary to probabilistically determine when traffic underneath the shaping is being generated at a faster or slower rate \cite{fu2003analytical}.

As an improvement to constant rate shaping, prior work suggests variable interval timer (VIT) shaping.  
VIT shaping uses a randomly varying send rate independent of the traffic that the shaping is intended to obfuscate.
If the varying rate has a high variance relative to the system jitter, it is resilient against analyses that defeat constant rate shaping \cite{fu2003analytical}. 
We performed a simulation of VIT on recorded packet captures from our tested devices. We found that VIT shaping adds no extra latency or bandwidth overhead relative to our constant-rate shaping implementation, because rate variations to overwhelm system jitter are orders of magnitude smaller than the total bandwidth overhead.  A real world deployment of traffic shaping, however, would have to mind this implementation detail.

\textbf{Shaping an entire smart home.}
A third-party hub or router deployment, like our traffic shaping implementation, would be able to shape all traffic from all devices in a smart home simultaneously.

Fortunately, shaping all of the traffic coming from a smart home is just as cost-effective as shaping traffic from a single device.
We have found that the data cost of ILP shaping does not scale linearly with the number of devices.  

In order to shape traffic from  multiple devices, you need a high enough shaped traffic rate to support the latency tolerances of all devices sending within a time window.
In practice, this rate is less than the sum of the rates that would be necessary to shape each device individually. 
Our simulation of VIT traffic shaping using a 24-hour packet capture from multiple smart home devices resulted in less cover traffic overhead than shaping the most bandwidth-intensive device individually. 
\section{Discussion}
\label{sec:discussion}

\textbf{Traffic shaping is practical.}
Our results indicate that independent link padding is reasonable given smart home Internet speeds and data caps. This is good news in light of past research that indicates that ILP shaping consumes too much extra data to be worthwhile \cite{dyer2012peek}. 

On the contrary, traffic shaping is well-suited to solving this particular smart home privacy problem.
The reasons for our success with ILP shaping are threefold:

First, smart home devices, especially non-A/V devices, use very little network bandwidth compared to smartphones and personal computers. 
This means that relatively small amounts of cover traffic are necessary to mask device activity. 
Smart home devices are also much more tolerant of long network latencies than devices with rich GUIs and interactive applications. 
This allows smart devices to properly function under slower network conditions imposed by ILP shaping.

Second, smart homes are likely to have Internet speeds and data caps much higher than typically used by consumers. For example, Comcast states that more than 99\% of consumers do not use their limit of 1TB of data per month \cite{comcast-usage}.  A constant small overhead due to traffic shaping will therefore have little to no effect on consumer experience.  Compare this with servers running anonymity network nodes -- a common deployment situation for traffic shaping -- that may always be at or near their upload or download speed limits. 

Third, ILP traffic shaping overheads do not scale linearly with the number of devices.  Adding more devices to a smart home does not necessarily increase the amount of shaping bandwidth needed. Although several smart home devices we tested do upload audio or video data and require higher traffic shaping rates, shaping at these rates is sufficient to also protect all other devices in the home. 

The cost amortization of ILP shaping also raises the possibility of shaping traffic elsewhere in the network to spread bandwidth and data costs over even more devices. Depending on the location of the adversary, traffic shaping could be performed at the level of an entire apartment complex, hotel, college campus, or neighborhood. In each of these cases, ILP traffic shaping could be performed on the router or routers connecting the protected community to the rest of the Internet.  As long as the adversary is upstream of the shaper and appropriate tunneling is used, no traffic rate metadata about any device in the protected community would be available to the adversary. Further research could explore the privacy protections and deployment issues of shaping IoT device traffic at large scale.

\textbf{Tunable traffic shaping.}
ILP traffic shaping could be implemented to allow a consumer to actively trade off privacy and data usage.  

For example, a user might know that no privacy sensitive activities will occur between midnight and 06:00. The traffic shaper could simply shut off during that period, using no data. Alternatively, the traffic shaper could reduce the shaped traffic rate to still protect lower-rate background traffic generated by devices while using less bandwidth overhead. Of course, an adversary would notice a change in the smart home traffic rate, but this is acceptable as long as no privacy sensitive activities (as determined by the user) are revealed. 

An actual consumer product could make these choices visible to the consumer. A smart home hub, for example, could allow a consumer to selectively choose which devices to protect via a smart phone interface. Or the interface could present the consumer with a privacy ``knob'' that allows them to tune their desired level of privacy, and use information from the ISP to calculate the actual financial cost of a certain privacy setting. In this way the actual privacy trade-off can be made real to the consumer. 

An even smarter hub-like device could learn from a user's activities and preferences and make an optimal tradeoff between privacy protection and cost. If the privacy properties of such algorithms can be rigorously defined and communicated in a user-friendly implementation, all smart home owners could benefit from privacy protections that adapt to their privacy values and Internet service limits.

\textbf{Options for device developers.}
Smart home device developers could implement devices with traffic shaping built-in, but there are more effective ways for developers of some devices to protect user privacy from traffic metadata analysis. 

Specifically, developers could implement devices such that network traffic doesn’t directly correspond to user behavior in real-time. A sleep monitor, for instance, does not need to update a cloud service immediately upon detecting that a user has gone to sleep---it could delay the notification for up to several hours without sacrificing functionality. Introducing long random delays in cloud notifications would also reduce an adversary’s confidence in inferred activities. 

However, devices that rely on real-time network communications are not amenable to these techniques. For example, a smart personal assistant cannot wait minutes to hours before answering a user’s questions. Developers could improve the privacy of these devices by making them more amenable to traffic shaping.

\textbf{Trust in VPNs.} Although our proposed solution of using traffic shaping with a VPN prevents a consumer's last-mile internet provider from performing device identification and behavior inference, the VPN exit node's last-mile provider could potentially perform the same attack.
This problem could be addressed by having the VPN provider act as an endpoint for traffic from multiple smart homes.
In that case, it would be difficult for the VPN's last-mile provider to associate devices with a particular smart home. 
We believe that such ``mixes'' for smart homes could work in a similar fashion as anonymity mixes in other contexts \cite{chaum1981untraceable, dingledine2004tor}.

\textbf{Minimum reliable product.} When testing our devices, we could clearly see that many manufacturers fail to include in the design of their device a strong ``minimum reliable product'' \cite{minimumreliableproduct}.
A smart device's ``minimum reliable product'' is its basic functionality in the absence of cloud support.
A strong minimum reliable product could allow a user to enjoy their smart device while blocking privacy-compromising network communications.
For example, a smart thermostat should, at the very least, match the functionality of a traditional thermostat if its server goes down.
Nest users in 2016 experienced what happens without a strong minimum reliable product when they lost heat in their homes due to Internet disconnections---an obviously poor user experience \cite{nestbroken}.
Policymakers and consumer advocacy groups should encourage device manufacturers to consider their device's minimum reliable product during development.

Policy measures have another role to play in encouraging the smart home ecosystem to protect consumers against the improper use of traffic rate metadata. 
Policymakers could regulate how actors like Internet service providers collect and use traffic rate information. 
The Federal Communications Commission's proposed rules in 2016 impacted how Internet Service Providers could collect and use data.
Though those rules were rolled back, we believe that future efforts should be sure to consider the impact of privacy regulations on smart homes.
\section{Conclusion}
\label{sec:conclusion}

The privacy threat of traffic metadata analysis will continue to grow along with the market for IoT smart home devices. 
In this paper, we demonstrate that a passive network adversary can infer  private in-home user activities from smart home traffic rates and packet headers \textit{even when devices use encryption}. 

We find that many commercially-available smart home devices do not function without network connectivity. This makes the smart home network metadata privacy problem unavoidable because these smart devices necessarily connect to the Internet. 

We show that traffic shaping by independent link padding can effectively protect smart home privacy. 
Despite commonly held beliefs, traffic shaping can be deployed in smart homes without significantly decreasing network performance or increasing data costs.

We hope that consumers will become better aware of smart home privacy vulnerabilities and incentivize manufacturers to more carefully consider privacy issues in the design of their products. 
While our proposed traffic shaping solution is practical and sufficient, improved regulation of ISPs and other passive network observers may also be necessary to offset the unique privacy challenges posed by IoT devices.

\bibliographystyle{abbrv}
\bibliography{SpyingOnTheSmartHome}

\end{document}